\documentclass{INTERSPEECH2023}

\interspeechcameraready

\usepackage{enumitem}
\usepackage{multirow}
\usepackage{makecell}
\usepackage{xhfill}
\usepackage{etoolbox}
\usepackage{tikz}
\usepackage{pgfplots}
\definecolor{plt_tabblue}{RGB}{31, 119, 180}
\definecolor{plt_tabyellow}{RGB}{255, 127, 14}
\definecolor{plt_tabgreen}{RGB}{44, 160, 44}
\definecolor{plt_tabgrey}{RGB}{127, 127, 127}
\usepgfplotslibrary{groupplots}
\usepgfplotslibrary{statistics}
\pgfplotsset{compat=1.16}
\makeatletter
\pgfplotsset{
  grid style={dotted, gray},
  boxplot/lower notch/.initial=\pgfplotsboxplotvalue{median},
  boxplot/upper notch/.initial=\pgfplotsboxplotvalue{median},
  boxplot/notch width/.initial=0.5,
  boxplot/draw/box/.code={%
    \draw[/pgfplots/boxplot/every box/.try]
      (boxplot box cs:\pgfplotsboxplotvalue{lower quartile},0)
      -- (boxplot box cs:\pgfplotsboxplotvalue{lower notch},0)
      -- (boxplot box cs:\pgfplotsboxplotvalue{median},0.5-\pgfplotsboxplotvalue{notch width}/2)
      -- (boxplot box cs:\pgfplotsboxplotvalue{upper notch},0)
      -- (boxplot box cs:\pgfplotsboxplotvalue{upper quartile},0)
      -- (boxplot box cs:\pgfplotsboxplotvalue{upper quartile},1)
      -- (boxplot box cs:\pgfplotsboxplotvalue{upper notch},1)
      -- (boxplot box cs:\pgfplotsboxplotvalue{median},0.5+\pgfplotsboxplotvalue{notch width}/2)
      -- (boxplot box cs:\pgfplotsboxplotvalue{lower notch},1)
      -- (boxplot box cs:\pgfplotsboxplotvalue{lower quartile},1)
      -- cycle
    ;
  },
  boxplot/draw/median/.code={%
    \draw[/pgfplots/boxplot/every median/.try]
        (boxplot box cs:\pgfplotsboxplotvalue{median},0.5-\pgfplotsboxplotvalue{notch width}/3*2)
        --
        (boxplot box cs:\pgfplotsboxplotvalue{median},0.5+\pgfplotsboxplotvalue{notch width}/3*2)
    ;
  }
}
\pgfplotsset{
  /pgfplots/custom legend/.style={
      legend image code/.code={
          \draw [|-|,#1] (0,2mm) -- node[rectangle,minimum size=2.5mm,draw,fill,##1]{}
          (0,7mm);
      }
  }
}

\title{Deep Multi-Frame Filtering for Hearing Aids}
\name{Hendrik Schröter$^1$, Tobias Rosenkranz$^2$, Alberto N. Escalante-B.$^2$, Andreas Maier$^1$}
\address{
  $^1$Friedrich-Alexander-Universit\"at Erlangen-N\"urnberg, Pattern Recognition Lab\\
  $^2$WS Audiology, Research and Development, Erlangen, Germany
}
\email{hendrik.m.schroeter@fau.de}

\newcommand{\R}{\mathbb{R}}
\newcommand{\C}{\mathbb{C}}

\DeclareMathOperator*{\argmin}{argmin}

\begin{document}

\maketitle
 
\begin{abstract}
  Multi-frame algorithms for single-channel speech enhancement are able to take advantage from short-time correlations within the speech signal.
  Deep filtering (DF) recently demonstrated its capabilities for low-latency scenarios like hearing aids with its complex multi-frame (MF) filter.
  Alternatively, the complex filter can be estimated via an MF minimum variance distortionless response (MVDR), or MF Wiener filter (WF).
  Previous studies have shown that incorporating algorithm domain knowledge using an MVDR filter might be beneficial compared to the direct filter estimation via DF.

  In this work, we compare the usage of various multi-frame filters such as DF, MF-MVDR, or MF-WF for HAs.
  We assess different covariance estimation methods for both MF-MVDR and MF-WF and objectively demonstrate an improved performance compared to direct DF estimation, significantly outperforming related work while improving the runtime performance.
\end{abstract}

\section{Introduction}

Hearing aids (HA) usually employ a filter bank~\cite{bauml2008uniform} similar to an STFT, as frequency transformation.
Subsequent processing steps, including single-channel noise reduction, is then performed in time/frequency (TF) domain.
The low-latency requirements of HAs of \SIrange{6}{10}{\ms}, however, usually result in a very poor frequency resolution.
This makes noise reduction within HAs particularly challenging since frequency resolution usually correlates well with noise reduction performance up to a certain point.
Especially single-frame Wiener filter approaches~\cite{hansler2005acoustic, aubreville2018deep, schroeter2020hcrnn} are used with a noise attenuation limit of \SIrange{6}{12}{\dB} since more attenuation would result in speech distortion and roughness.
This is because a one-tap Wiener filter reduces to a single real-valued gain and thus is not able to recover the clean phase.
Other options, like a complex ratio mask (CRM), are able to theoretically restore the original phase.
However, especially in low-latency scenarios the available frequency resolution may be limited down to \SI{250}{\Hz}.
For a low fundamental frequency, this may result in up to 5 speech harmonics within one frequency bin which makes estimating a phase correction factor inherently harder for the CRM~\cite{schroeter2022lowlatency}.
Therefore, complex filters \cite{mack2019deep, schroeter2020clcnet} introduced as deep filtering (DF) have been used for allowing a stronger noise attenuation in HAs~\cite{schroeter2022lowlatency}.
Further, DF outperformes complex ratio masks, especially with a low frequency resolution of HA filter banks \cite{schroeter2022deepfilternet, schroeter2022lowlatency}

Recently, deep MF beamforming filters have been proposed in contrast to direct estimation of the filter coefficients within DF~\cite{huang2011multi, xu2020neural, tammen2021deep, zhang2021multi}.
In contrast to classical beamforming using multiple channels, the inter-frame correlations are used to derive a complex filter in TF domain.
Huang~\cite{huang2011multi} proposed to decompose multi-frame speech signal into a inter-frame correlated component and a interfering component.
This assumption allowed them to introduce an MF-MVDR beamformer with classical parameter estimation.
Zhang et al.~\cite{zhang2021multi} proposed to use DF to estimate a clean speech signal which is then used for classical estimation of the MVDR parameters.
Similarly, Pan et.al~\cite{pan2022dnn} used a CRM followed by an estimation of MF Wiener or MVDR filter statistics.
However, MF MVDR and Wiener filter perform worse compared to the only CRM enhanced output signal e.g.~in terms of PESQ.
Tamen et al.~\cite{tammen2021deep} proposed a deep MF-MVDR filter where a neural network was used to estimate the inter-frame correlation matrices of speech and noise signals.
The authors reported that the deep MF-MVDR filter outperforms direct DF estimation.

In this work we follow \cite{tammen2021deep} and estimate the covariance matrices directly using a DNN.
We evaluate different covariance estimation methods and compare DF to deep multi-frame MVDR and Wiener filters.

\section{Multi-Frame Filtering}
\label{sec:mf}
\subsection{Signal Model}
\label{ssec:signalmodel}
Let $x(k)$ be a mixture signal
\vspace{-.25em}\begin{equation}
  x(k) = s(k) + z(k)\text{,}
  \label{eq:sigmodel}
\end{equation}
where $s(k)$ is a clean speech signal and $z(k)$ an interfering background noise.
Typically, HA noise reduction operates in time/frequency domain:
\begin{equation}
  X(t, f) = S(t, f) + Z(t, f)\text{,}
\end{equation}
where $X(t, f)$ is the filter bank representation of the time domain signal $x(k)$ and $t$, $f$ are the time and frequency bins.

\subsection{Deep Filtering and Multi-Frame Signal Model}
\label{ssec:df}

Deep filtering was proposed to take advantage from short-time correlations within the speech signal \cite{schroeter2020clcnet} and for signal reconstruction e.g.~by destructive interference or package loss \cite{mack2019deep}.
In the following, we describe the initial deep filter proposal, where the filter weights are directly estimated by a deep neural network (DNN).
Further, we make the bridge to multi-frame processing using MVDR or Wiener filtering and describe the MF signal model taking advantage of speech inter-frame correlations.

Deep filtering is defined by a complex filter in TF-domain \cite{mack2019deep, schroeter2020clcnet}:
\begin{equation}
  Y(t, f) = \sum_{i=0}^{N-1} W^{*}_i(t, f) \cdot X(t - i + l, f)\text{\ ,}
  \label{eq:DF}
\end{equation}
where $W^{*}$ are the complex conjugated coefficients of filter order $N$ that are applied to the input spectrogram $X$, and $\hat{Y}$ the enhanced spectrogram.
$l$ is an optional look-ahead, which allows incorporating non-causal taps in the linear combination if $l\ge1$.
In previous work, additionally also included filtering along the frequency axis allowing to incorporate correlations e.g.~due to overlapping bands~\cite{schroeter2022lowlatency}, which is not considered in this study.
This of course could also be used within the beamforming algorithms.

To simplify the following, we omit the frequency index $f$ since all frequency bins are processed equivalently.
Further, with filter length $N$, we define the noisy multi-frame vector as $\bm{\bar{x}}_{t} \in \C^N$:
\begin{equation}
  \bm{\bar{x}}(t)= [X(t + l), X(t - 1 + l), \dots, X(t -N+1 +l)]^\text{T}\text{\ .}
\end{equation}
And with the complex filter $\bm{\bar{w}}(t) \in \C^N$
\begin{equation}
  \bm{\bar{w}}(t)= [W_0(t), W_1(t), \dots, W_{N-1}(t)]^\text{T}
\end{equation}
the complex filter of Equation~(\ref{eq:DF}) reduces to:
\begin{equation}
  \boxed{Y(t) = \bm{\bar{w}}_{\text{DF}}(t)^{\text{H}}\bm{\bar{x}}(t)}\text{\ ,}
  \label{eq:df2}
\end{equation}
where $\circ^\text{H}$ denotes the conjugate transpose operator.
As mentioned above, deep filtering directly estimates the complex filter $\bm{\bar w}_\text{DF}(t)$.
However, $\bm{\bar w}(t)$ can also be estimated using multi-frame beamforming algorithms which will be described in the following.

Assuming speech and noise are uncorrelated (which is requirement for Eq.~\ref{eq:sigmodel}), the noisy covariance matrix $\bm{\Phi}_{yy}(t)\in\C^{N\times N}$ is given by
\begin{equation}
  \bm{\Phi}_{yy}(t) = E[\bm{\bar y}(t) \bm{\bar y}^\text{H}(t)] =  \bm{\Phi}_{ss}(t) +  \bm{\Phi}_{zz}(t)\text{,}
  \label{eq:cov}
\end{equation}
where $E[\circ]$ is the mathematical expectation.
The matrices $\bm{\Phi}_{ss}(t)$ and $\bm{\Phi}_{zz}(t)$ are defined analogously.

We further assume after \cite{huang2011multi, benesty2011single} that the speech signal consists of a \textit{desired}, short-time correlated component $\bm{\bar s}^c$ and an uncorrelated, interfering component $\bm{\bar s}^i$ wrt.~the speech coefficient $S(t)$:
\begin{equation}
  \bm{\bar s}(t) = \bm{\bar s}^c(t) + \bm{\bar s}^i(t)
  \label{eq:s_cor1}
\end{equation}
with 
\begin{equation}
  \bm{\bar s}^c(t) = \bm{\bar\gamma}_{s}(t) S(t)\text{.}
  \label{eq:s_cor2}
\end{equation}
The speech inter-frame correlation (IFC) vector $\bm{\bar\gamma}_{s}(t)$ is highly time-varying and is defined as
\begin{equation}
  \bm{\bar\gamma}_{s}(t) = \dfrac{E[\bm{s}(t) S(t)^*]}{E[|S(t)|^2]} = \dfrac{\bm{\Phi}_{ss} e}{e^\text{T}\bm{\Phi}_{ss} e}\text{\ ,}
  \label{eq:speech_ifc}
\end{equation}
where $e=[1, 0, 0, \dots, 0]^\text{T}\in\R^{N}$ is the $N$-dimensional selection vector.
Note, that a different selection index may be used e.g.~when using non-causal taps within the filter.
The denominator $e^\text{T}\bm{\Phi}_{ss} e$ corresponds to the speech power spectral density (PSD) $\phi_s(t)$.
Thus, the first element of the speech IFC vector equals 1:
\begin{equation}
  e^{\text{T}}\bm{\bar\gamma}_{s}(t) = 1\text{.}
  \label{eq:ifc_1}
\end{equation}
When considering the uncorrelated speech component as interference, with (\ref{eq:s_cor1}) and (\ref{eq:s_cor2}), the \textit{multi-frame signal model} is given by
\begin{equation}
  \boxed{\bm{\bar x}(t) = \bm{\bar\gamma}_s(t)S(t) + \bm{\bar u}(t)}\text{\ ,}
  \label{eq:mf_signalmodel}
\end{equation}
where $\bm{\bar u}(t) = \bm{\bar s}^i(t) + \bm{\bar z}(t)$ is the undesired noise and interference vector.

\subsection{Multi-Frame Wiener Filter}
\label{ssec:wf}

As mentioned above, single-frame (tap) Wiener filters reduce to a single real-valued gain.
In the following we describe the general form resulting in a complex filter $\bm{\bar{w}}(t)$.

The Wiener filter tries to directly minimize the difference between clean speech $S(t)$ and the prediction $Y(t)$ using the mean squared error (MSE):
\begin{equation}
  \begin{split}
    \bm{\bar w}_{\text{WF}}(t) &= \argmin_{\bm{\bar w}} E[|S(t)-Y(t)|^2] \\ 
    & = \argmin_{\bm{\bar w}} E[|S(t)-\bm{\bar w}^\text{H}(t)\bm{\bar x}(t)|^2] 
  \end{split}
  \label{eq:wf_mse}
\end{equation}

With the uncorrelation assumption between speech and noise, the solution of (\ref{eq:wf_mse}) is given by
\begin{equation}
  \boxed{\bm{\bar w}_\text{WF}(t) = \bm\Phi_{xx}^{-1} E[\bm{\bar x}(t)S(t)]=\bm\Phi_{xx}^{-1}{\bm{\bar\gamma}_s}}\text{\ .}
  \label{eq:wf}
\end{equation}

\subsection{Multi-Frame MVDR Filter}
\label{ssec:mvdr}

In contrast to Wiener filtering which tries to be optimal wrt.~SNR, the MVDR filter is optimal wrt.~speech distortion.
Given a standard filter-and-sum beamformer~\cite{doclo2015multichannel}
\begin{equation}
  E[|Y(t)|^2] = E[\bm{\bar w}^\text{H}  \bm{\bar x}\bm{\bar x}^\text{H} \bm{\bar w}] = \bm{\bar w}\bm\Phi_{xx}\bm{\bar w}^\text{H}\text{,}
  \label{eq:filtersum}
\end{equation}
the following distortionless response constraint requires that the predicted output $Y(t)$ is equal to the target speech $S(t)$:
\begin{equation}
  Y(t) = \bm{\bar w}^\text{H}(t) \bm{\bar \gamma}_s(t) S(t) \overset{!}{=} S(t)
  \label{eq:mvdr_constraint}
\end{equation}
Now, the MVDR filter can be defined as
\begin{equation}
  \min_{\bm{\bar w}} \bm{\bar w}^\text{H}(t)\bm\Phi_{xx}(t)\bm{\bar w}(t)\text{, s.t.~}\bm{\bar w}^\text{H}(t)\bm{\bar \gamma}_s(t)=1\text{.}
  \label{eq:mvdr_min}
\end{equation}

Solving this minimization problem leads to the MF-MVDR beamformer \cite{capon1969high, van1988beamforming, huang2011multi}:
\begin{equation}
  {\bm{\bar w}_{\text{MVDR}}(t) = \dfrac{\bm\Phi_{xx}^{-1}(t)\bm{\bar\gamma}_{s}}{\bm{\bar\gamma}_{s}^\text{H}\bm\Phi_{xx}^{-1}\bm{\bar\gamma}_{s}}}\text{\ .}
  \label{eq:mvdr}
\end{equation}
Following \cite{van1988beamforming, doclo2015multichannel}, we assume noise $\bm{\bar{u}}(t)$ and $\bm{\bar{s}}(t)$ being uncorrelated, we can rewrite $\bm\Phi_{xx}$ using (\ref{eq:mf_signalmodel})
\begin{equation}
  \vspace{-.2em}
  \bm\Phi_{xx}(t) = \phi_s(t) \bm{\bar\gamma}_{s}(t) \bm{\bar\gamma}_{s}^\text{H}(t) + \bm\Phi_{uu}(t)\text{,}
  \label{eq:phixx}
\end{equation}
where $\bm\Phi_{uu}(t)$ represents the undesired noise and interference covariance matrix.
With (\ref{eq:phixx}), it can be shown \cite{van1988beamforming, doclo2015multichannel} that the MVDR beamformer can be rewritten to
\begin{equation}
  \vspace{-.2em}
  \boxed{\bm{\bar w}_{\text{MVDR}}(t) = \dfrac{\bm\Phi_{uu}^{-1}(t)\bm{\bar\gamma}_{s}}{\bm{\bar\gamma}_{s}^\text{H}\bm\Phi_{uu}^{-1}\bm{\bar\gamma}_{s}}}\text{\ .}
  \label{eq:mvdr2}
\end{equation}

\subsection{Filter Estimation}
\label{ssec:filter_est}

To estimate filter weights $\bm{\bar w}^\text{WF}$ and $\bm{\bar w}^\text{MVDR}$ we need to estimate the speech IFC vector $\bm{\bar\gamma}_{s}$ as well as the covariance matrices $\bm\Phi_{xx}$ or $\bm\Phi_{uu}$.
Similar to \cite{tammen2022deep}, we directly estimate the IFC vector $\bm{\bar\gamma}_{s}\in\C^{N}$ followed by a normalization to fulfill (\ref{eq:ifc_1}).

Within preliminary experiments, we discovered that estimating the noisy covariance matrix $\bm\Phi_{xx}(t)$ using a DNN provides better results compared to estimating it via statistics like~\cite{pan2022dnn}.
We explain this with the update speed of noisy covariance matrix and IFC vector.
A DNN estimate is superior over a recursive update of $\bm\Phi_{xx}(t)$~\cite{pan2022dnn} since it can adapt the update speed depending on the current noise and speech conditions.
Second, we estimate the noise covariance matrix in Eq.~(\ref{eq:mvdr2}) \cite{souden2009optimal, fischer2018robust, tammen2021deep}, in contrast to \cite{pan2022dnn} who used the MVDR implementation of Eq.~(\ref{eq:mvdr}).\\[.5em]
We compare the following configurations for covariance estimation:
\begin{enumerate}[noitemsep,topsep=0pt,parsep=0pt,partopsep=0pt]
  \item \textbf{Direct estimation.} We directly estimate $\bm\Phi_{xx}$, and $\bm\Phi_{uu}$. For numerical stability, we apply diagonal loading of \num{1e-7} before matrix inversion.
  \item \textbf{Inverse estimation.} To avoid computing the inverse of $\bm\Phi$, we directly estimate the inverse $\bm\Phi^{-1}$.
  \item \textbf{Hermitian.} As stated above, the covariance matrix can be assumed to be Hermitian positive-definite. Thus, we can define the Hermitian PSD $\bm H(t)$ via
    \vspace{-.2em}
    \begin{equation}
      \bm\Phi(t) = \bm{H}(t)\bm{H}^\text{H}(t)\text{,}
      \label{eq:hermitian}
      \vspace{-.2em}
    \end{equation}
    where $\bm H(t)\in\C^{N\times N}$ is the Hermitian matrix~\cite{tammen2021deep}.
    By estimating $\bm H$, the matrix multiplication ensures that the resulting covariance matrices fulfill its hermitian properties.
  \item \textbf{Hermitian of inverse.}  Since the inverse $\bm\Phi^{-1}$ is also Hermitian positive-definite, we estimate the Hermitian PSD of the inverse:
  \vspace{-.2em}
  \begin{equation}
      \bm\Phi^{-1}(t) = \bm{H}(t)\bm{H}^\text{H}(t)\text{.}
      \label{eq:hermitian_i}
      \vspace{-.2em}
    \end{equation}
\end{enumerate}

We further tested enforcing Hermitian properties of the predicted covariance or estimating a Cholesky decomposition of the predicted Hermitian similar to~\cite{tammen2022deep}.
However, the results presented in section~\ref{ssec:prelim_cov} did not change significantly.


\section{Training Framework}

\subsection{DeepFilterNet Framework}

We adopt the perceptual approach of DeepFilterNet~\cite{schroeter2022deepfilternet, schroeter2022deepfilternet2} which also has been used for hearing aids~\cite{schroeter2022lowlatency}.
The two-stage denoising process takes advantage of auditory properties which allows for relatively efficient DNN.
That is, the first stage only operates in real-valued ERB (equivalent rectangular bandwidth) domain and tries to recover the speech envelope.
The second stage uses MF filtering to enhance the periodic part of speech up to a frequency of $f_\text{mf}=\SI{4}{\kHz}$ which covers most of the energy of the periodic speech component.

Instead of a short-time Fourier transformation (STFT) used in~\cite{schroeter2022deepfilternet2}, we employ a \SI{24}{kHz} uniform polyphase hearing aid filter bank~\cite{bauml2008uniform} with 48 frequency bands and a frequency resolution of \SI{250}{\Hz}.
The filter bank roughly corresponds to an STFT with a window size of \SI{4}{\ms} and a hop size of \SI{1}{\ms}.
Note that other filter banks~\cite{lollmann2005generalized} may achieve a better frequency analysis resolution which was not considered since we want to be able to integrate this method into an existing hearing aid setup.

We apply both denoising stages in parallel for practical reasons like better concurrency possibilities.
Hence, the MF filter is applied to the noisy spectrum instead of the pre-enhanced spectrum of stage 1 unlike in~\cite{schroeter2022deepfilternet2}.
Further, for the MVDR and Wiener filter estimation we add another grouped linear output layer for the covariance matrix estimation.
Even though the DNN input and output is complex-valued, the DNN only operates on real-valued tensors.
The full source code except the hearing aid filter bank is publicly available at \url{https://github.com/rikorose/deepfilternet}.

\begin{figure}
  \includegraphics[width=\linewidth]{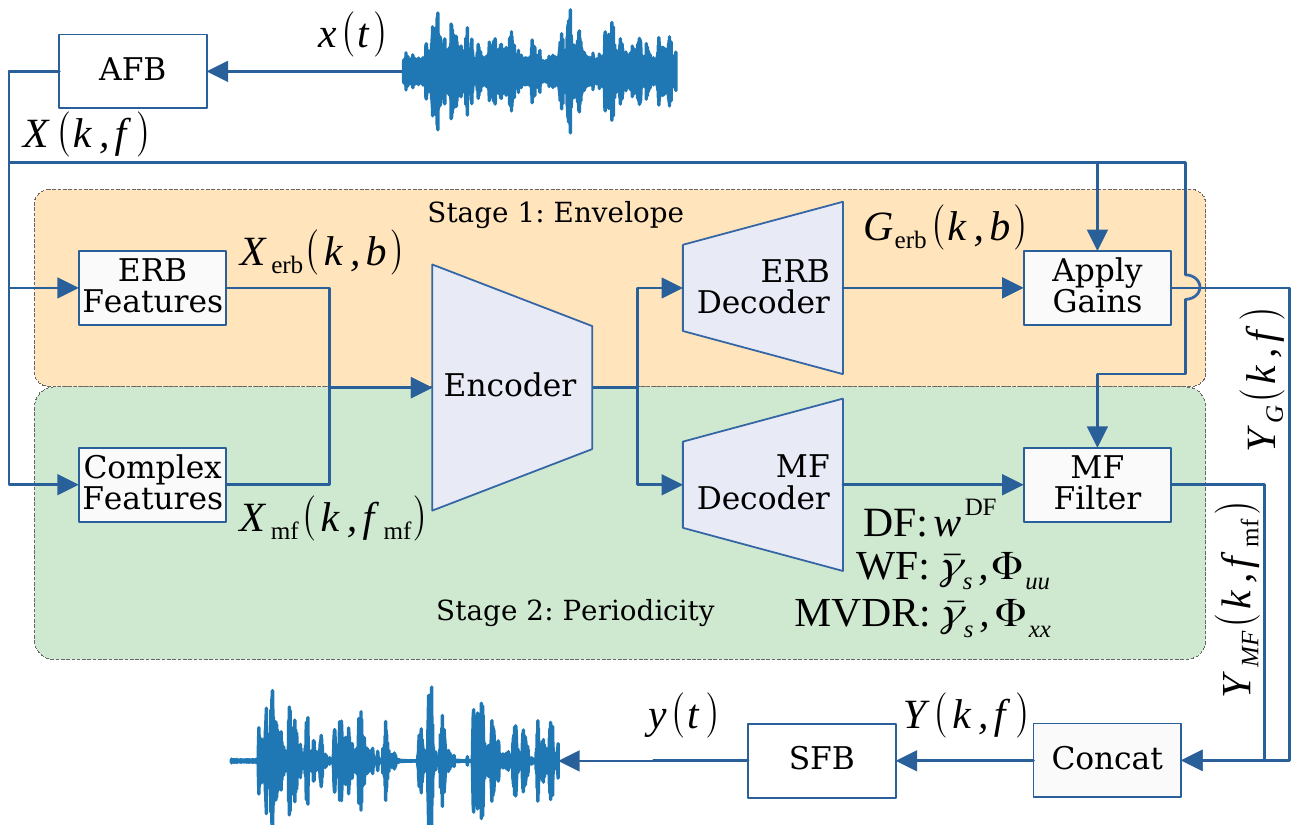}
  \caption{Two-stage noise reduction framework using a real-valued ERB stage followed by the multi-frame (MF) filtering stage based on~\cite{schroeter2022deepfilternet2}.
    Instead of (I)STFT, we employ hearing aid analysis (AFB) and synthesis filter banks (SFB).
    Depending on the configuration the second stage predicts directly the deep filter coefficient $\bm w^\text{DF}$, or the speech IFC vector $\bm{\bar\gamma}_{s}$ and covariance matrices $\bm\Phi$ for the Wiener and MVDR filters.}
  \label{fig:framework}
  \vspace{-1.5em}
\end{figure}

\vspace{-0.2em}\subsection{Datasets and Training}
\vspace{-0.2em}

We use the multi-lingual DNS4 dataset~\cite{dubey2022icassp} for training.
Following~\cite{schroeter2022deepfilternet}, we oversample the included high quality datasets VCTK~\cite{} and PTDB by a factor of 10.
Moreover, we trimmed silences and filtered the dataset with DNSMOS V4~\cite{dubey2022icassp} to only include samples with overall mean opinion score (OVRL MOS) greater than 3.
We split the datasets into training/development/test (70/15/15$\%$).
VCTK and PTDB were split on speaker level ensuring no overlap with the VCTK/Demand test set~\cite{valentini2016investigating}.
The remaining English read speech and noise datasets are split on signal level.
We conduct preliminary experiments using only VCTK + PTDB as speech datasets and report results on the VCTK/Demand test set \cite{valentini2016investigating}.
For evaluation of the final models, we further report results on the recent DNS5 track 2 blind test set~\cite{dubey2022icassp} and an internal test set recorded with HAs containing 30 noisy samples without groundtruth.

Data preprocessing and augmentation is adopted from \cite{schroeter2022deepfilternet2}.
Additionally, we resample the data to \SI{24}{\kHz} to match the filter bank sampling rate.
Declipping and dereverberation were not considered in this work.

We decreased the model size by reducing the convolution channels to 16 and the number of hidden units of the GRU layers to 128 resulting in \SI{510}{k} parameters of the DNN.
We adopted the loss function from~\cite{schroeter2022deepfilternet2}, trained all models for 100 epochs, and applied early stopping based on the development set.
We use AdamW optimizer, an initial learning rate of 0.001, learning rate decay of 0.5 per epoch, and weight decay of 0.05.
The latter is especially important for stable gradient due to complex number processing, matrix inversions and division with small numbers e.g.~in (Eq. \ref{eq:mvdr2}).

\section{Experiments}

The following performance metrics were employed to evaluate our multi-frame filtering approaches.
The time-domain scale-invariant signal-distortion-ratio (SI-SDR)~\cite{le2019sdr}, and the frequency domain measures PESQ~\cite{rix2001PESQ} and STOI~\cite{taal2011stoi}, as well as the composite measures CSIG, CBAK and COVL~\cite{hu2007evaluation}. Further, we adopt the ``pseudo''-subjective measure DNSMOS V5~\cite{dubey2022icassp} for judging signal quality of signals without groundtruth.
Further, we provide the real-time-factor (RTF) on a notebook Core-i5 quad-core CPU for inference speed evaluation.

\vspace{-0.5em}\subsection{Covariance Estimation}%
\vspace{-0.2em}\label{ssec:prelim_cov}

We conducted preliminary experiments to find the best way of estimating the covariance matrices $\bm\Phi_{xx}(t)$ (WF) and $\bm\Phi_{uu}(t)$ (MVDR).
As we can see in Table~\ref{tab:prelim_wo_constraints}, estimating the Hermitian of $\bm\Phi$ or $\bm\Phi^{-1}$ seems crucial for the MVDR filter as $\bm\Phi_{uu}(t)$ is not invertible with the direct estimate.
Directly estimating the inverse covariance matrix, resulted in a distorted audio which did not improve during training.
The Wiener filter however, is not so sensitive wrt.~the different estimation methods as only small differences can be observed.
The estimated Hermitian matrix provides a noticeable benefit for the MF Wiener filter.
Thus, for further experiments, we chose to estimate the Hermitian PSD of the covariance inverse for both WF and MVDR (i.e.~row 4 and 8 of Table~\ref{tab:prelim_wo_constraints}).


\begin{table}
  \caption{Comparison of different covariance estimation options of Section~\ref{ssec:filter_est} based on the VCTK/Demand dataset. ``Invert.'' stands for estimating the inverse covariance matrix, ``Herm.'' stands for estimating $\bm H(t)$ instead of $\bm\Phi$ as in Equations~(\ref{eq:hermitian}), (\ref{eq:hermitian_i}). Bold values denote best results for this metric.}
  \label{tab:prelim_wo_constraints}
  \robustify\bfseries
  \sisetup{
    table-number-alignment = center,
    table-figures-integer  = 1,
    table-figures-decimal  = 2,
    table-auto-round = true,
    detect-weight = true
  }
  \addtolength{\tabcolsep}{-4pt}
  \resizebox{\linewidth}{!} {
    \begin{tabular}{lccSSS[table-figures-decimal=3]SSS}
    \toprule
    \multicolumn{1}{c}{\rotatebox[origin=c]{60}{Filter}} & \multicolumn{1}{c}{\rotatebox[origin=c]{60}{Invert.}} & \multicolumn{1}{c}{\rotatebox[origin=c]{60}{Herm.}} & \multicolumn{1}{c}{\rotatebox[origin=c]{60}{SI-SDR}} & \multicolumn{1}{c}{\rotatebox[origin=c]{60}{PESQ}} & \multicolumn{1}{c}{\rotatebox[origin=c]{60}{STOI}} & \multicolumn{1}{c}{\rotatebox[origin=c]{60}{CSIG}} & \multicolumn{1}{c}{\rotatebox[origin=c]{60}{CBAK}} & \multicolumn{1}{c}{\rotatebox[origin=c]{60}{COVL}} \\
    \midrule
    MF-WF &  &  & 17.00 & 2.61 & 0.924 & 3.66 & 3.23 & 3.13\\
    MF-WF &  & \checkmark & 17.10 & 2.62 & 0.925 & 3.65 & 3.22 & 3.11\\
    MF-WF & \checkmark &  & 17.06 & 2.62 & 0.925 & 3.61 & 3.24 & 3.11\\
    MF-WF & \checkmark & \checkmark & 17.13 & 2.63 & 0.926 & 3.69 & 3.23 & 3.15\\
    MF-MVDR &  & & \multicolumn{6}{c}{\makecell{------------ $\bm\Phi_{uu}$ not invertible ------------}}\\
    MF-MVDR &  & \checkmark & 16.81 & 2.53 & 0.921 & 3.61 & 3.17 & 3.06 \\
    MF-MVDR & \checkmark & & \multicolumn{6}{c}{\makecell{------------- Bad convergence -------------}}\\
    MF-MVDR & \checkmark & \checkmark & \bfseries 17.31 & \bfseries 2.65 & \bfseries 0.929 & \bfseries 3.70 & \bfseries 3.24 & \bfseries 3.17 \\
    \bottomrule
  \end{tabular}
  \vspace{-.6em}
  }
\end{table}

\vspace{-0.5em}\subsection{Comparison of MF Deep Filtering / Wiener Filter / MVDR Filter}

We evaluated our models on the VCTK/Demand test set and provide a comparison with related algorithms using a hearing aid filter bank with the same frequency resolution\cite{aubreville2018deep, schroeter2022lowlatency}.
As we can see in table~\ref{tab:resutls_vbdemand}, our all proposed multi-frame filters outperform related work. 
Further, MF-WF and MF-MVDR provide slightly superior results over direct filter estimation using deep filtering.

\begin{table}
  \caption{Objective results on the VCTK/Demand dataset.
  All models use a uniform polyphase filter bank and introduce an algorithmic latency of 8~ms including 2 frames of look-ahead.
  Number of parameter (Params) in million.}
  \label{tab:resutls_vbdemand}
  \robustify\bfseries
  \sisetup{
    table-number-alignment = center,
    table-figures-integer  = 1,
    table-figures-decimal  = 2,
    table-auto-round = true,
    detect-weight = true
  }
  \addtolength{\tabcolsep}{-3pt}
  \resizebox{\linewidth}{!} {
    \begin{tabular}{lS[table-figures-decimal=2]SSSS[table-figures-decimal=3]SSS}
    \toprule
      \multicolumn{1}{c}{\rotatebox[origin=c]{60}{Filter}} & \multicolumn{1}{c}{\rotatebox[origin=c]{60}{Params}} & \multicolumn{1}{c}{\rotatebox[origin=c]{60}{RTF}} & \multicolumn{1}{c}{\rotatebox[origin=c]{60}{SI-SDR}} & \multicolumn{1}{c}{\rotatebox[origin=c]{60}{PESQ}} & \multicolumn{1}{c}{\rotatebox[origin=c]{60}{STOI}} & \multicolumn{1}{c}{\rotatebox[origin=c]{60}{CSIG}} & \multicolumn{1}{c}{\rotatebox[origin=c]{60}{CBAK}} & \multicolumn{1}{c}{\rotatebox[origin=c]{60}{COVL}} \\
    \midrule
    WF \cite{aubreville2018deep} & 50.00 & \multicolumn{1}{c}{-} & 8.94 & 2.12 & 0.942 & 3.33 & 2.43 & 2.79 \\
    MF-DF \cite{schroeter2022lowlatency} & 0.87 & 0.25 & 14.04 & 2.65 & 0.938 & 4.01 & 3.17 & 3.32 \\
    MF-DF & \bfseries 0.51 & \bfseries 0.18 & \bfseries 18.21 & 2.85 & 0.943 & 4.09 & 3.39 & 3.46 \\
    MF-WF & 0.53 & 0.19 & 17.94 & \bfseries 2.91 & 0.943 & \bfseries 4.13 & 3.42 & \bfseries 3.52 \\
    MF-MVDR & 0.53 & 0.19 & 18.18 & 2.90 & \bfseries 0.945 & 4.12 & \bfseries 3.43 & 3.51 \\
    \bottomrule
  \end{tabular}
  }
\end{table}

\begin{figure}
  \centering
    \begin{tikzpicture}
  \begin{axis}[
    ymin=1.0,
    ymax=4.5,
    y tick label style={/pgf/number format/.cd, fixed, fixed zerofill, precision=1, /tikz/.cd},
    ylabel={MOS},
    x tick label as interval,
    xtick={0, 1, 2, 3, 4, 5, 6, 7, 8, 9, 10, 11, 12},
    xticklabels={SIG, BAK, OVRL},
    grid=major,
    cycle list={{plt_tabblue},{plt_tabyellow},{plt_tabgreen},{plt_tabgrey}},
    boxplot={,
      draw direction=y,
      box extend=0.2,
      every median/.style={ultra thick},
      draw position={1/5 + floor(\plotnumofactualtype/4) + 1/5*mod(\plotnumofactualtype,4)},
    },
    width=1.0\linewidth,
    height=0.5\linewidth,
    boxplotcolor/.style={color=#1,mark options={color=#1,mark size=1.5pt}},
    legend entries = {Noisy, DF, WF, MVDR},
    legend style={at={(0.5,-0.25)}, anchor=north,legend columns=-1, font=\footnotesize},
    legend image code/.code={
      \draw[#1, draw=none] (0cm,-0.1cm) rectangle (0.6cm,0.1cm);
    },
  ]
    \addplot+[
        mark=*,
        boxplotcolor=plt_tabblue,
        fill, fill opacity=0.5,
        boxplot prepared={
            median=3.2216182975732135,
            lower quartile=2.7898907510366127,
            upper quartile=3.4257300480784343,
            lower whisker=1.8921941100436,
            upper whisker=3.693724868122197,
        }]
        coordinates {
            (1.0, 1.2796167021636615) (1.0, 1.6230331538063256) (1.0, 1.204126841795205) (1.0, 1.8357078440105188) (1.0, 1.6798516850276524) (1.0, 1.418716815637935) (1.0, 1.2622969426686708) (1.0, 1.1738797724045358) (1.0, 1.573349209913116) (1.0, 1.333569585370992) (1.0, 1.1625532920909816) (1.0, 1.2663191576954045) (1.0, 1.207395396169726) (1.0, 1.8059833415341504) (1.0, 1.661861361971335) (1.0, 1.6558788383884224) (1.0, 1.684542292310392) (1.0, 1.2137962107681182) (1.0, 1.165249261753936) (1.0, 1.3147182332639076) (1.0, 1.2985827274857171) (1.0, 1.238204197152645) (1.0, 1.311429641889659) (1.0, 1.175120036823355) (1.0, 1.7935506305781097) (1.0, 1.4565366296555384) (1.0, 1.333882897959279) (1.0, 1.2725547361915832) (1.0, 1.170643232715254) (1.0, 1.202678808888146) (1.0, 1.6175713686018731) (1.0, 1.268283397549401) 
        };
    \addplot+[
        mark=*,
        boxplotcolor=plt_tabyellow,
        fill, fill opacity=0.5,
        boxplot prepared={
            median=3.0966137931148188,
            lower quartile=2.858783134249043,
            upper quartile=3.3426530003525814,
            lower whisker=2.1386688264836504,
            upper whisker=3.696885134425382,
        }]
        coordinates {
            (1.0, 2.089136545238274) (1.0, 2.1184828314591475) (1.0, 1.875626691853962) (1.0, 2.0015284837650427) (1.0, 2.034503697383832) (1.0, 1.9980709434103952) (1.0, 1.9921638577535703) (1.0, 1.757049331582325) (1.0, 2.1204004107081977) (1.0, 2.1032379027324155) 
        };
    \addplot+[
        mark=*,
        boxplotcolor=plt_tabgreen,
        fill, fill opacity=0.5,
        boxplot prepared={
            median=3.0787770773741143,
            lower quartile=2.846420568867627,
            upper quartile=3.360686723207011,
            lower whisker=2.0967016214933323,
            upper whisker=3.699846923509652,
        }]
        coordinates {
            (1.0, 1.950266197274592) (1.0, 1.815248968006171) (1.0, 1.9204150481977524) (1.0, 1.663746909759767) (1.0, 1.937876135433214) (1.0, 2.055909028188531) (1.0, 1.823169986830006) 
        };
    \addplot+[
        mark=*,
        boxplotcolor=plt_tabgrey,
        fill, fill opacity=0.5,
        boxplot prepared={
            median=3.0659735780882085,
            lower quartile=2.8505069551763773,
            upper quartile=3.346467337472604,
            lower whisker=2.115498631570252,
            upper whisker=3.710385133315594,
        }]
        coordinates {
            (1.0, 2.05598779956822) (1.0, 2.106028855491397) (1.0, 1.925940241319196) (1.0, 2.072358163128788) (1.0, 1.8194938985202087) (1.0, 1.992129100796213) (1.0, 1.883566912069359) (1.0, 1.97656436181953) 
        };
    \addplot+[
        mark=*,
        boxplotcolor=plt_tabblue,
        fill, fill opacity=0.5,
        boxplot prepared={
            median=2.445582984743459,
            lower quartile=1.86693911105238,
            upper quartile=2.939972723674152,
            lower whisker=1.0869260932121343,
            upper whisker=4.156634011193534,
        }]
        coordinates {
            
        };
    \addplot+[
        mark=*,
        boxplotcolor=plt_tabyellow,
        fill, fill opacity=0.5,
        boxplot prepared={
            median=3.623915932877088,
            lower quartile=3.1518035490500154,
            upper quartile=3.8494404862278877,
            lower whisker=2.139400867982692,
            upper whisker=4.216822196552725,
        }]
        coordinates {
            (1.0, 1.8076982262051364) (1.0, 2.0447650900034824) (1.0, 1.953504282404262) (1.0, 2.0731833200234626) (1.0, 1.8213244778491475) (1.0, 1.6551030038330907) (1.0, 2.097306218957887) (1.0, 1.770549754271142) (1.0, 1.892149974364606) (1.0, 2.0752477624420997) 
        };
    \addplot+[
        mark=*,
        boxplotcolor=plt_tabgreen,
        fill, fill opacity=0.5,
        boxplot prepared={
            median=3.698942683110607,
            lower quartile=3.240206178295514,
            upper quartile=3.9047050150081484,
            lower whisker=2.258515067554684,
            upper whisker=4.214637790795832,
        }]
        coordinates {
            (1.0, 1.9945190468830656) (1.0, 2.211732760539464) (1.0, 2.0687296547846348) (1.0, 1.8015900572560068) (1.0, 2.143720537120416) (1.0, 2.236785209910906) (1.0, 1.951647086346069) (1.0, 2.126642073308237) (1.0, 2.201407050595769) 
        };
    \addplot+[
        mark=*,
        boxplotcolor=plt_tabgrey,
        fill, fill opacity=0.5,
        boxplot prepared={
            median=3.6277442845668038,
            lower quartile=3.135475275697136,
            upper quartile=3.8464325734077898,
            lower whisker=2.069302776121875,
            upper whisker=4.204837471185156,
        }]
        coordinates {
            (1.0, 1.9583070307700703) (1.0, 1.8325807602697852) (1.0, 1.9877043704044817) (1.0, 1.775067283668104) (1.0, 2.004130151311537) (1.0, 1.894513941255844) (1.0, 2.0638961595113354) (1.0, 1.9881459665890304) (1.0, 1.9162074009076835) (1.0, 1.940147530002035) (1.0, 2.0025287666899776) (1.0, 2.0492054192949203) (1.0, 2.041058918598122) (1.0, 1.9655078168421367) (1.0, 1.719906767625052) (1.0, 1.5784854645477793) 
        };
    \addplot+[
        mark=*,
        boxplotcolor=plt_tabblue,
        fill, fill opacity=0.5,
        boxplot prepared={
            median=2.2009910875320156,
            lower quartile=1.8414137967723732,
            upper quartile=2.514559153122214,
            lower whisker=1.0680324560753132,
            upper whisker=3.378841139680811,
        }]
        coordinates {
            
        };
    \addplot+[
        mark=*,
        boxplotcolor=plt_tabyellow,
        fill, fill opacity=0.5,
        boxplot prepared={
            median=2.5598994503110015,
            lower quartile=2.30227349238053,
            upper quartile=2.870569735910309,
            lower whisker=1.4906179705175724,
            upper whisker=3.489733356956517,
        }]
        coordinates {
            (1.0, 1.3888657550046288) (1.0, 1.4196676881570307) 
        };
    \addplot+[
        mark=*,
        boxplotcolor=plt_tabgreen,
        fill, fill opacity=0.5,
        boxplot prepared={
            median=2.564925456768073,
            lower quartile=2.3171301621758813,
            upper quartile=2.8828399657156942,
            lower whisker=1.5533337022478664,
            upper whisker=3.516568249929515,
        }]
        coordinates {
            (1.0, 1.375380866276009) 
        };
    \addplot+[
        mark=*,
        boxplotcolor=plt_tabgrey,
        fill, fill opacity=0.5,
        boxplot prepared={
            median=2.548025040501686,
            lower quartile=2.2897125115004737,
            upper quartile=2.848525921701512,
            lower whisker=1.5114271853120034,
            upper whisker=3.5167745132230324,
        }]
        coordinates {
            (1.0, 1.3318778930199828) 
        };
  \end{axis}
\end{tikzpicture}
  \caption{DNSMOS V5 \cite{dubey2022icassp} on the DNS5 blind test set.}
  \label{fig:dnsmos_dns5}
\end{figure}
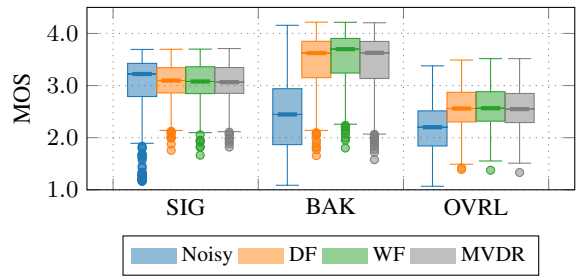

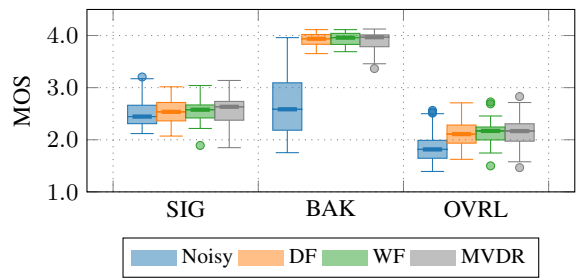
\begin{figure}
  \centering
    \begin{tikzpicture}
  \begin{axis}[
    ymin=1.0,
    ymax=4.5,
    y tick label style={/pgf/number format/.cd, fixed, fixed zerofill, precision=1, /tikz/.cd},
    ylabel={MOS},
    x tick label as interval,
    xtick={0, 1, 2, 3, 4, 5, 6, 7, 8, 9, 10, 11, 12},
    xticklabels={SIG, BAK, OVRL},
    grid=major,
    cycle list={{plt_tabgrey},{plt_tabblue},{plt_tabyellow},{plt_tabgreen}},
    boxplot={,
      draw direction=y,
      box extend=0.2,
      every median/.style={ultra thick},
      draw position={1/5 + floor(\plotnumofactualtype/4) + 1/5*mod(\plotnumofactualtype,4)},
    },
    width=1.0\linewidth,
    height=0.5\linewidth,
    boxplotcolor/.style={color=#1,mark options={color=#1,mark size=1.5pt}},
    legend entries = {Noisy, DF, WF, MVDR},
    legend style={at={(0.5,-0.25)}, anchor=north,legend columns=-1, font=\footnotesize},
    legend image code/.code={
      \draw[#1, draw=none] (0cm,-0.1cm) rectangle (0.6cm,0.1cm);
    },
  ]
    \addplot+[
        mark=*,
        boxplotcolor=plt_tabblue,
        fill, fill opacity=0.5,
        boxplot prepared={
            median=2.444950392635002,
            lower quartile=2.308802916740864,
            upper quartile=2.6621593550227596,
            lower whisker=2.1187355634502496,
            upper whisker=3.171329928983547,
        }]
        coordinates {
            (1.0, 3.204130754638096) 
        };
    \addplot+[
        mark=*,
        boxplotcolor=plt_tabyellow,
        fill, fill opacity=0.5,
        boxplot prepared={
            median=2.5357554905336954,
            lower quartile=2.3620516446650717,
            upper quartile=2.7155122500120976,
            lower whisker=2.070625374353306,
            upper whisker=3.0161639470505635,
        }]
        coordinates {
            
        };
    \addplot+[
        mark=*,
        boxplotcolor=plt_tabgreen,
        fill, fill opacity=0.5,
        boxplot prepared={
            median=2.576508046172238,
            lower quartile=2.418390895783733,
            upper quartile=2.6677891703966163,
            lower whisker=2.216714056867184,
            upper whisker=3.040989566324053,
        }]
        coordinates {
            (1.0, 1.8911952713956708) 
        };
    \addplot+[
        mark=*,
        boxplotcolor=plt_tabgrey,
        fill, fill opacity=0.5,
        boxplot prepared={
            median=2.6326176117588673,
            lower quartile=2.3750657382425944,
            upper quartile=2.7362236726849893,
            lower whisker=1.8485626381921192,
            upper whisker=3.137688425745556,
        }]
        coordinates {
            
        };
    \addplot+[
        mark=*,
        boxplotcolor=plt_tabblue,
        fill, fill opacity=0.5,
        boxplot prepared={
            median=2.5844925362853015,
            lower quartile=2.1832095324208636,
            upper quartile=3.092480991701332,
            lower whisker=1.7522681137267924,
            upper whisker=3.959516178526166,
        }]
        coordinates {
            
        };
    \addplot+[
        mark=*,
        boxplotcolor=plt_tabyellow,
        fill, fill opacity=0.5,
        boxplot prepared={
            median=3.937945208979211,
            lower quartile=3.8319024139865294,
            upper quartile=4.020129595506586,
            lower whisker=3.65308848739901,
            upper whisker=4.113578145787317,
        }]
        coordinates {
            
        };
    \addplot+[
        mark=*,
        boxplotcolor=plt_tabgreen,
        fill, fill opacity=0.5,
        boxplot prepared={
            median=3.959086585753564,
            lower quartile=3.8290300747973056,
            upper quartile=4.041078390746839,
            lower whisker=3.690196173690662,
            upper whisker=4.113412781583385,
        }]
        coordinates {
            
        };
    \addplot+[
        mark=*,
        boxplotcolor=plt_tabgrey,
        fill, fill opacity=0.5,
        boxplot prepared={
            median=3.967252751554353,
            lower quartile=3.785178681219939,
            upper quartile=4.011870742245324,
            lower whisker=3.458429568399491,
            upper whisker=4.124872620348845,
        }]
        coordinates {
            (1.0, 3.366289522189092) 
        };
    \addplot+[
        mark=*,
        boxplotcolor=plt_tabblue,
        fill, fill opacity=0.5,
        boxplot prepared={
            median=1.8150179320789053,
            lower quartile=1.6451541200430089,
            upper quartile=1.9888574355737083,
            lower whisker=1.3918263775426334,
            upper whisker=2.4999851367156323,
        }]
        coordinates {
            (1.0, 2.5348729991227) (1.0, 2.5105928166210343) (1.0, 2.5628157109638776) 
        };
    \addplot+[
        mark=*,
        boxplotcolor=plt_tabyellow,
        fill, fill opacity=0.5,
        boxplot prepared={
            median=2.108206804597172,
            lower quartile=1.9348743590529138,
            upper quartile=2.282263228961433,
            lower whisker=1.624461073038473,
            upper whisker=2.7084816459513776,
        }]
        coordinates {
            
        };
    \addplot+[
        mark=*,
        boxplotcolor=plt_tabgreen,
        fill, fill opacity=0.5,
        boxplot prepared={
            median=2.1696556788541894,
            lower quartile=1.9964054857796178,
            upper quartile=2.244579874623118,
            lower whisker=1.7456398632342405,
            upper whisker=2.456065120055672,
        }]
        coordinates {
            (1.0, 1.4992189646195606) (1.0, 2.725548543988634) (1.0, 2.6890975322760275) 
        };
    \addplot+[
        mark=*,
        boxplotcolor=plt_tabgrey,
        fill, fill opacity=0.5,
        boxplot prepared={
            median=2.1666796022622186,
            lower quartile=1.9725449858129875,
            upper quartile=2.3063335313335642,
            lower whisker=1.5776603535299203,
            upper whisker=2.714546761163052,
        }]
        coordinates {
            (1.0, 1.4651698785236058) (1.0, 2.8289138990609177) 
        };
  \end{axis}
\end{tikzpicture}
  \caption{DNSMOS V5 \cite{dubey2022icassp} on the internal HA test set.}
  \label{fig:dnsmos_internal_ha}
\end{figure}

Figures~\ref{fig:dnsmos_dns5} and~\ref{fig:dnsmos_internal_ha} provide ``pseudo''-subjective measures on two noisy datasets without a groundtruth.
On the DNS5 blind test set, MF-WF achieves the highest background MOS (BAK), that is, the lowest background distortion.
The MF-MVDR model however, is able to retain more speech compared to DF and WF within the internal HA dataset.

\begin{figure}[tbh]
  \includegraphics[width=\linewidth,trim=1.2cm 0.6cm 2cm 0.5cm,clip]{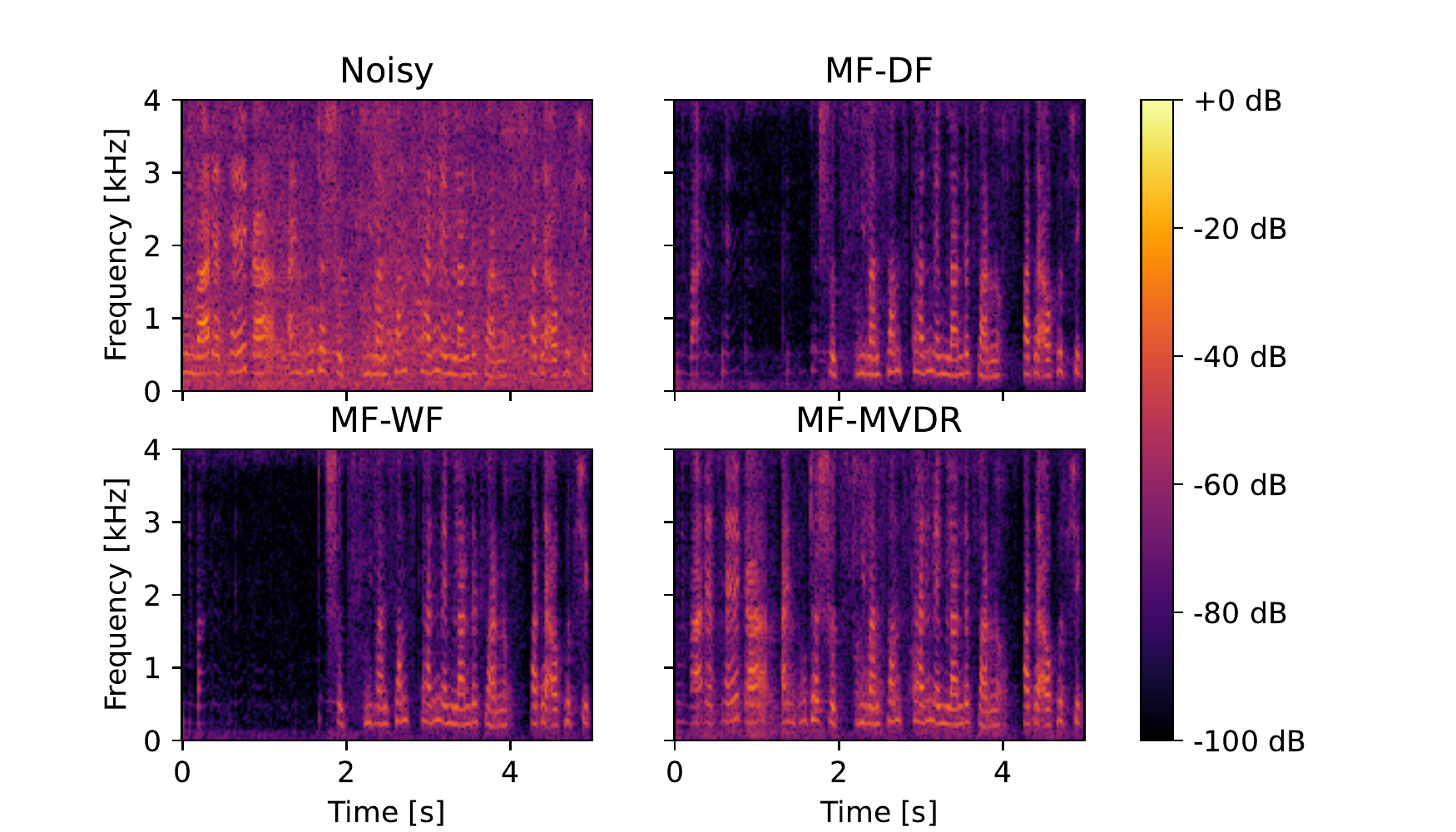}
  \caption{Sample from the internal HA test set.}%
  \label{fig:spectrograms}
  \vspace{-1em}
\end{figure}

This can further be observed in a qualitative figure of a sample from the HA test set (Figure~\ref{fig:spectrograms}).
While DF and similarly also MF-WF wrongly suppresses the first few seconds of speech in a noisy HA recording, the MF-MVDR is able to quickly adopt to new speech.
This matches with the motivation of both multi-frame filters.
The MF-WF provides a stronger noise suppression at the cost of more speech degradation, the MF-MVDR filter however, tries to keep speech distortion at a minimum level and sacrifices a little noise reduction in return.

\section{Conclusions}

In this study, we presented a deep learning-based multi-frame filtering method for hearing aids.
We evaluated different methods of estimating the covariance matrices for MF-WF and MF-MVDR and provided evidence that the presented MF Wiener filter and MVDR filter outperform direct filter estimation.

Especially the MF-MVDR filter is relevant for hearing aid usage, since minimal speech distortion is one of the key requirements of HA noise reduction algorithms.
Although noise is suppressed robustly, when running  separate instances on the left and right hearing aids we observed spatial distortions specifically with MF-DF and MF-WF processing.
Therefore, further research is needed in the area of filter synchronization between devices.

\bibliographystyle{IEEEtran}
\bibliography{refs}

\end{document}